\documentclass{iau379}

\usepackage{mathrsfs}
\usepackage{amsfonts}
\usepackage{amstext}
\usepackage{amsmath}
\usepackage{graphics,graphicx}
\usepackage{multirow}

\newcommand{\hi } {{\rm H}\,{\small\rm I}}

\begin{document}

\lefttitle{Federico Lelli}
\righttitle{Mass models of disk galaxies from gas dynamics}

\jnlPage{1}{7}
\jnlDoiYr{2023}
\doival{10.1017/xxxxx}

\aopheadtitle{Proceedings of IAU Symposium 379}
\editors{P. Bonifacio,  M.-R. Cioni, F. Hammer, M. Pawlowski, and S. Taibi, eds.}

\title{Mass models of disk galaxies from gas dynamics}

\author{Federico Lelli$^1$}
\affiliation{$^1$ INAF - Arcetri Astrophysical Observatory, Largo Enrico Fermi 5, 50125, Florence, Italy}

\begin{abstract}
I review methods and techniques to build mass models of disk galaxies from gas dynamics. I focus on two key steps: (1) the derivation of rotation curves using 3D emission-line datacubes from \hi, CO, and/or H$\alpha$ observations, and (2) the calculation of the gravitational field from near-infrared images and emission-line maps, tracing the stellar and gas mass distributions, respectively. Mass models of nearby galaxies led to the establishment of the radial acceleration relation (RAR): the observed centripetal acceleration from rotation curves closely correlates with that predicted from the baryonic distribution at each galaxy radius, even when dark matter supposedly dominates the gravitational field. I conclude by discussing the (uncertain) location of Local Group dwarf spheroidal galaxies on the RAR defined by more massive disk galaxies.
\end{abstract}

\begin{keywords}
Dark matter --- galaxies: kinematics and dynamics --- galaxies: spiral --- galaxies: dwarf
\end{keywords}

\maketitle

\vspace{-0.5cm}
\section{Introduction}

Rotation curves from gas kinematics are one of the most direct methods to measure the gravitational potential ($\Phi$) of galaxies and, therefore, constrain their baryonic and dark matter (DM) content. The reason is simple. In gas disks, the rotation velocity ($V_{\rm rot}$) is typically much larger than the velocity dispersion ($\sigma_{\rm V}$), so pressure support is negligible ($V_{\rm rot}/\sigma_{\rm V} \gg 1)$ and $V_{\rm rot}$ is a direct proxy of the circular velocity of a test particle ($V_{c}^2 = -R\,\nabla \Phi$). For example, this is not the case for the stellar components of galaxies because both rotation and pressure support can be important, so $V_{\rm c}$ must be estimated by solving the Jeans' equations, which require either full 6D phase-space information (as for the Milky Way thanks to the Gaia mission) or assumptions on the velocity dispersion tensor (as for external galaxies with stellar spectroscopy).

Over the past 20 years, the study of gas dynamics made enormous progress thanks to dedicated surveys of different emission lines, probing different gas phases: the \hi\ line at 21 cm tracing atomic gas, CO lines in the submm tracing molecular gas, and Balmer lines in the optical (mostly the H$\alpha$ line) tracing warm ionized gas. Another key advancement has been the availability of near infrared (NIR) images from space telescopes, such as Spitzer and WISE. The NIR light is only marginally affected by dust extinction and represents the best tracer of the stellar mass distribution, so it allows us to accurately compute the stellar gravitational field.

For galaxies in groups and in the field, \hi\ disks are typically more extended than the stellar components (on average by $\sim$4 times the NIR half-light radii $R_{\rm h}$, \citealt{lelli2016a}), so they trace the gravitational potential out to the most DM-dominated regions. To date, the main disadvantage of \hi\ observations is the low spatial resolution (typically $5''-30''$) but the situation will drastically improve with the upcoming Square Kilometer Array (SKA). On the other hand, CO and H$\alpha$ disks are typically confined to the inner galaxy regions, but can be routinely observed at spatial resolutions of $\sim1''-2''$, which can be pushed further down to sub-arcsec resolutions thanks to adaptive optics for H$\alpha$ data and ALMA long baselines for CO data. The best approach to study gas kinematics is a multiwavelength one, combing CO and/or H$\alpha$ data at high resolution in the inner parts with \hi\ data at lower resolutions in the outer regions.

In this brief review, I focus on methods and techniques to measure circular-velocity curves (Sect. \ref{sec:rotation}) and build mass models of galaxies (Sect. \ref{sec:massmodels}). The scientific implications of these observational and modeling efforts have been extensively reviewed elsewhere \citep{deBlok2010, McGaugh2020, Lelli2022} and will not be repeated here. I will only mention the radial acceleration relation (RAR) in context with Local Group (LG) dwarf galaxies (Sect. \ref{sec:rar}).

\section{Measuring the circular-velocity curve}\label{sec:rotation}

\subsection{Circular motions}

For a geometrically and optically thin disk with nearly circular orbits, the projected line-of-sight velocity $V_{\rm l.o.s.}$ at a sky position $(x, y)$ is given by
\begin{equation}\label{eq:Vlos}
 V_{\rm los}(x, y) = V_{\rm sys} + V_{\rm rot}(R)\sin(i)\cos(\theta),
\end{equation}
where $V_{\rm sys}$ is the systemic velocity (due to the Hubble expansion and peculiar motions), $V_{\rm rot}$ is the rotational velocity at radius $R$ in the galaxy plane, $i$ is the inclination angle between the normal to the disk and the line of sight, and $\theta$ is the azimuthal angle in the disk plane given by
\begin{equation}\label{eq:theta}
 \cos(\theta) = \dfrac{-(x - x_{0})\sin(P.A.) + (y - y_{0})\cos(P.A.)}{R},
\end{equation}
where ($x_0$, $y_0$) are the center coordinates and $P.A.$ is the position angle taken in anti-clockwise direction between the North direction and the major axis of the projected disk. Thus, the rotation curve $V_{\rm rot}(R)$ can be measured knowing six parameters: $V_{\rm sys}$, $x_0, y_0$, $P.A.$, and $i$.

The most basic approach to measure $V_{\rm rot}$ is to obtain slit spectroscopy along the disk major axis ($\theta=0^{\circ}$), so that $V_{\rm rot}(R) = [V_{\rm los}(R)-V_{\rm sys}]/\sin(i)$. This 1D approach must assume $(x_0, y_0$), $P.A.$, and $i$ from independent observations (e.g., optical images) and cannot account for warped disks in which $P.A.$ and $i$ can vay with $R$. A better strategy is to obtain velocity maps ($V_{\rm l.o.s.}$ at each sky position) and fit them with Eq.\,\ref{eq:Vlos} \citep{Begeman1989}. This 2D approach is robust for well-resolved disks with hundreds of resolution elements, but becomes unreliable in less resolved disks (e.g., dwarf and/or distant galaxies) because of beam-smearing effects, which systematically underestimate $V_{\rm rot}$ and overestimate $\sigma_{\rm V}$ \citep[e.g.,][]{DiTeodoro2015}.

The best approach is to fit directly the 3D emission-line cube (with two spatial axes and one velocity axis) to reproduce the full shape of the line profiles at each $(x, y)$. In this 3D approach, disk models are built starting from some rotation curve $V_{\rm rot}(R)$, intrinsic velocity dispersion profile $\sigma_{\rm V}(R)$, gas surface density profile $\Sigma_{\rm gas}(R)$, and disk thickness $z_{\rm gas}$. The disk models are projected on the sky to produce a mock cube, which is smoothed to the same spectral and spatial resolution of the data (reproducing beam smearing effects) and compared with the observed cube until a best-fit is found \citep{Swaters2009}. The 3D approach exploits the full information available in the data, but requires more free parameters than the 2D approach to model $\sigma_{\rm V}$, $\Sigma_{\rm gas}$, and $z_{\rm gas}$. In most cases, $\Sigma_{\rm gas}$ can be inferred from the observed intensity map (with the exception of edge-on galaxies), while $z_{\rm gas}$ is generally unmeasurable because is below the angular resolution, so it is fixed to sensible values. Over the past years, several public 3D fitting softwares became available: $^{\rm 3D}$Barolo \citep{DiTeodoro2015} and FAT \citep{Kamphuis2015} adopt a so-called \emph{tilted-ring modeling}, while KinMS \citep{Davis2013} and GalPak3D \citep{Bouche2015} use a \emph{parametric modeling}.

The tilted-ring modeling \citep{Rogstad1974, Begeman1989} divides the gas disk into $N$ rings (whose separation is typically set by the spatial resolution) and fits Eq.\,\ref{eq:Vlos} independently in each ring. It is common to perform multiple, iterative fits. After a first fit, $V_{\rm sys}$ and ($x_0$, $y_0$) are measured as the mean (or median) values across the rings and fixed in subsequent fits with less free parameters. The angles $P.A.$ and $i$ can be estimated in the same way unless the gas disk is warped: in such cases the radial variations of $P.A.$ and $i$ are fitted with appropriate smooth functions, which are then imposed in a final fit to infer $V_{\rm rot}$ and $\sigma_{\rm V}$ in each ring.

The parametric modeling \citep[e.g.,][]{Courteau1997} uses parameteric functions to describe $V_{\rm rot}(R)$ and fits a global model to the whole disk. Warps can be accounted for, but require a ``trial and fail'' procedure to choose appropriate functions to model $P.A.(R)$ and $i(R)$. The parametric modeling has less free parameters than the tilted-ring modeling; this facilitates the use of Markov Chain Monte Carlo algorithms \citep{Bouche2015} and neural networks \citep{Dawson2021}. In the parametric modeling, $V_{\rm rot}$ may be described by an empirical fitting function, such as $V_{\rm t}\arctan(R/R_{\rm t})$ \citep{Courteau1997}, or by a mass model that includes different velocity components (see Sect.\,\ref{sec:massmodels}). The parametric modeling, however, does not provide an actual, empirical derivation of the rotation curve because a smooth shape is imposed, neglecting possible real features in $V_{\rm rot}(R)$. Moreover the intrinsic $\sigma_{\rm V}$ is often described by a single, radially averaged value, which may be dominated by the inner bright regions during the 3D fit. Alternatively, one must assume a parametric functional form for $\sigma_{\rm V}(R)$ as well.

%In this situation, the derivation of velocity maps is ambiguous due to beam-smearing effects that alter the intrinsic shape of the line profiles \citep{Swaters2002a}. Gas emission with different intrinsic $V_{\rm l.o.s}$ is flux-averaged within the resolution element, so the line profiles tend to become broad and non-Gaussian with long tails of emission towards the galaxy systemic velocity. The latter effect occurs because of the $\cos(\theta)$ term in Eq.\,\ref{eq:Vlos} that reduces velocity projections far from the major axis. In such cases, if one estimates $V_{\rm l.o.s}$ as the first moment along the profile or as the central velocity of a Gaussian fit, the rotational velocity will be underestimated \citep{Swaters2009, Lelli2010}. Moreover, if one estimates the gas velocity dispersion $\sigma_{\rm V}$ as the second moment along the profile or as the width of a Gaussian fit, the intrinsic velocity dispersion will be severely overestimated \citep{DiTeodoro2015}. Both effects are especially strong in the inner regions because there is a maximal crowding of different velocity projections. There is no simple way to correct for beam smearing using a 2D approach because it depends on the interplay between intrinsic rotation-curve shape, gas distribution, disk thickness, inclination, angular and spectral resolutions.

\subsection{Noncircular motions}\label{sec:noncircular}

Noncircular motions may affect the measured $V_{\rm rot}\simeq V_{\rm c}$. The simplest form of non-circular motion is a radial flow ($V_{\rm rad}$) in the disk plane, giving an additional term $V_{\rm rad}(R)\sin(i)\sin(\theta)$ in the right-end side of Eq.\,\ref{eq:Vlos}. More generally, $V_{\rm rot}$ and $V_{\rm rad}$ are the first-order terms of an harmonic expansion of the line-of-sight velocity \citep{Schoen1997}:
\begin{equation}
  V_{\rm l.o.s.}(x, y) = V_{\rm sys} + \sin(i)\sum_{m=1} \left [ c_{m}(R)\cos(m\theta) + s_{m}(R)\sin (m\theta) \right ],
\end{equation}
where the harmonic order $m=1$ gives $c_{1} = V_{\rm rot}$ and $s_{1} = V_{\rm rad}$. A perturbation of the gravitational potential of order $m$ causes harmonics of order $m-1$ and $m+1$ in projection \citep{Schoen1997}. For example, bar-like and oval distortions with $m=2$ give $m=1$ and $m=3$ terms in $V_{\rm l.o.s}$ \citep{Spekkens2007}. The harmonic decomposition further increases the number of free parameters in tilted-ring fits, so it can only be applied to high-resolution, high-sensitivity data. To date, this method has been applied up to $m=3$ with 2D fits, but further progress may be done using a 3D approach that models the full line profiles.

\hi\ studies find that noncircular motions are typically smaller than 10 km\,s$^{-1}$, corresponding to $\sim$1$\%$ to $\sim$10$\%$ of $V_{\rm rot}$ \citep{Gentile2005, Trachternach2008, Oh2015, Marasco2018}. They may be larger in barred and starburst galaxies, and become progressively more important in dwarf galaxies with small $V_{\rm rot}$ \citep[e.g.,][]{Lelli2012b}. Nevertheless, nonciruclar motions are usually comparable to uncertainties in $V_{\rm rot}$, which are computed considering the difference in $V_{\rm rot}$ from the approaching and receding sides of the disk \citep{Swaters2009}. These uncertainties are not formal errors, but quantify global kinematic asymmetries, effectively capturing noncircular motions in most cases. Noncircular motions could be larger in H$\alpha$ disks \citep{Simon2005, Spekkens2007}, possibly because ionized gas is more closely related to the sites of star formation activity than atomic gas. %Importantly, observed noncircular motions are not strong enough to ``hide'' a central DM cusp \cite{Gentile2005, Trachternach2008, Oh2015}. Opposite conclusions are found analysing simulated dwarf galaxies \cite{Oman2019} because non-circular motions in simulated dwarfs appears to be stronger than in real ones \cite{Marasco2018}.

\subsection{Random motions}\label{sec:pressure}

Pressure support due to random motions is often negligible in gas disks. For atomic and molecular gas, the intrinsic $\sigma_{\rm V}$ is typically between $5-15$ km\,s$^{-1}$, so effectively $V_{\rm c} = V_{\rm rot}$ for galaxies with $V_{\rm rot}\gtrsim 50$ km\,s$^{-1}$ \citep{Swaters2009} and pressure support become important only in the tiniest gas-bearing dwarfs \citep{Iorio2017}. The situation can be different for ionized gas because the intrinsic $\sigma_{\rm V}$ can be higher ($\sim15-30$ km\,s$^{-1}$), so pressure support can be important in galaxies with $V_{\rm rot}\lesssim 100$ km\,s$^{-1}$ \citep{Barat2020}. For both atomic and ionized gas, the velocity dispersion is largely driven by turbulent gas motions rather than thermal motions, so the pressure support is also referred to as turbulence support.

For gas disks with $V_{\rm rot}/\sigma_{\rm V}\lesssim4$, pressure support can be accounted for using the so-called asymmetric-drift correction (ADC, \citealt{BT94}). Assuming that $\sigma_{\rm V}$ is isotropic (as expected for collisional gas due to frequent energy exchanges), we obtain
\begin{equation}\label{eq:drift1}
V_{\rm c}^2 = V_{\rm rot}^2 - \sigma_{\rm V}^2 \left( \dfrac{\partial \ln \rho_{\rm gas}}{\partial \ln R} + \dfrac{\partial \ln \sigma_{\rm V}^2}{\partial \ln R} \right),
\end{equation}
where $\rho_{\rm gas}(R,z)$ is the gas volume density, which is not directly observable. 

A common approach \citep{Meurer1996} is to assume that the vertical density distribution does not vary with radius, so $\partial \ln \rho_{\rm gas} / \partial \ln R = \partial \ln \Sigma_{\rm gas} / \partial \ln R$, where the gas surface density profile $\Sigma_{\rm gas}(R)$ can be traced from emission-line maps. Then, one may either estimate the radial derivatives of $\Sigma_{\rm gas}$ and $\sigma_{\rm V}$ separately, or fit the product $\Sigma_{\rm gas} \sigma_{\rm V}^2$ with a smooth parametric function with a trivial radial derivative \citep{Iorio2017}. The main caveat with this approach is that the thickness of \hi\ disks may actually increase with radius \citep{Bacchini2020}.

Another approach to solve Eq.\,\ref{eq:drift1} \citep{Burkert2010} is to consider a self-gravitating disk where $\sigma_{\rm V}$ is independent of $z$, so $\rho_{\rm gas}(R, z) = \rho_{0}(R) {\rm sech}^2(z/z_{\rm d})$ \citep{Spitzer1942} where $\rho_{0}(R) = \pi G_{\rm N} \Sigma_{\rm gas}^2(R)/2 \sigma_{\rm V}^2(R)$ is the volume density at $z=0$ and $G_{\rm N}$ is Newton's constant. The gas disks of galaxies, however, are not self-gravitating. A self-consistent ADC should use an iterative approach: (i) measure $\Phi(R, z)$ fitting some initial $V_{\rm c}(R)$ with a 3D mass model (Sect.\,\ref{sec:massmodels}), (ii) measure the vertical structure as a function of $R$ assuming hydrostatic equilibrium for the given $\Phi(R, z)$, (iii) measure again $V_{\rm rot}$, $\sigma_{\rm V}$, and $V_{\rm c}$ for the new vertical structure, and (iv) iterate.

At any rate, the dominant uncertainty in the ADC is driven by $\sigma_{\rm V}(R)$. In the relevant cases (dwarf galaxies), $\sigma_{\rm V}$ is often poorly measured and assumed to be constant with radius. Then, for an exponential gas disk with scale length $R_{\rm d}$, Eq.\,\ref{eq:drift1} simplifies to $V_c^2 =V_{\rm rot}^2 + \sigma_{\rm V}^2(R/R_{\rm d})$. This equation can be used as a zeroth-order approximation in poorly resolved galaxies.

\section{Mass Models}\label{sec:massmodels}

\subsection{Measuring the baryonic gravitational field}\label{sec:math}

To study the relative distribution of baryons and DM in galaxies from the measured $V_{\rm c}(R)$, it is necessary to compute the Newtonian gravitational acceleration from various mass components \citep{vanAlbada1986}. In cylindrical coordinates $(R, z)$, a razor-thin exponential disk \citep{Freeman1970} gives the following velocity contribution in the disk plane ($z=0$):
\begin{equation}\label{eq:Freeman}
 V_{\rm c}^2(R) = \dfrac{G_{\rm N} M_{\rm d}}{R_{\rm d}} 2 y^{2} \left[ I_0(y)K_0(y) - I_1(y)K_1(y) \right]
\end{equation}
where $M_{\rm d}$ the disk mass, $R_{\rm d}$ the disk scale length, $y=R/(2R_{\rm d})$, and $I_n$ and $K_n$ are modified Bessel functions of the first and second kind, respectively. A razor-thin exponetial disk is just a zeroth-order representation of the mass distribution in galaxies because it neglects the finite thickness of stellar and gas disks, as well as relevant features in their mass distribution such as inner concentrations (bulges, pseudobulges, nuclei), inner depressions, outer breaks, bumps, wiggles, and so on. For a more realistic model, we need to numerically solve $\nabla^2 \Phi_{i} = 4\pi G_{\rm N} \rho_{i}$ for each baryonic component $i$. There are two main approaches: (1) to measure the projected radial density profile $\Sigma(R)$ and make assumptions on the intrinsic 3D geometry, or (2) to fit 2D images with the multi-Gaussian expansion (MGE) method \citep{Emsellem1994}.

\begin{figure}[t]
\begin{center}
  \centerline{\vbox to 6pc{\hbox to 10pc{}}}
  \includegraphics[width=\textwidth]{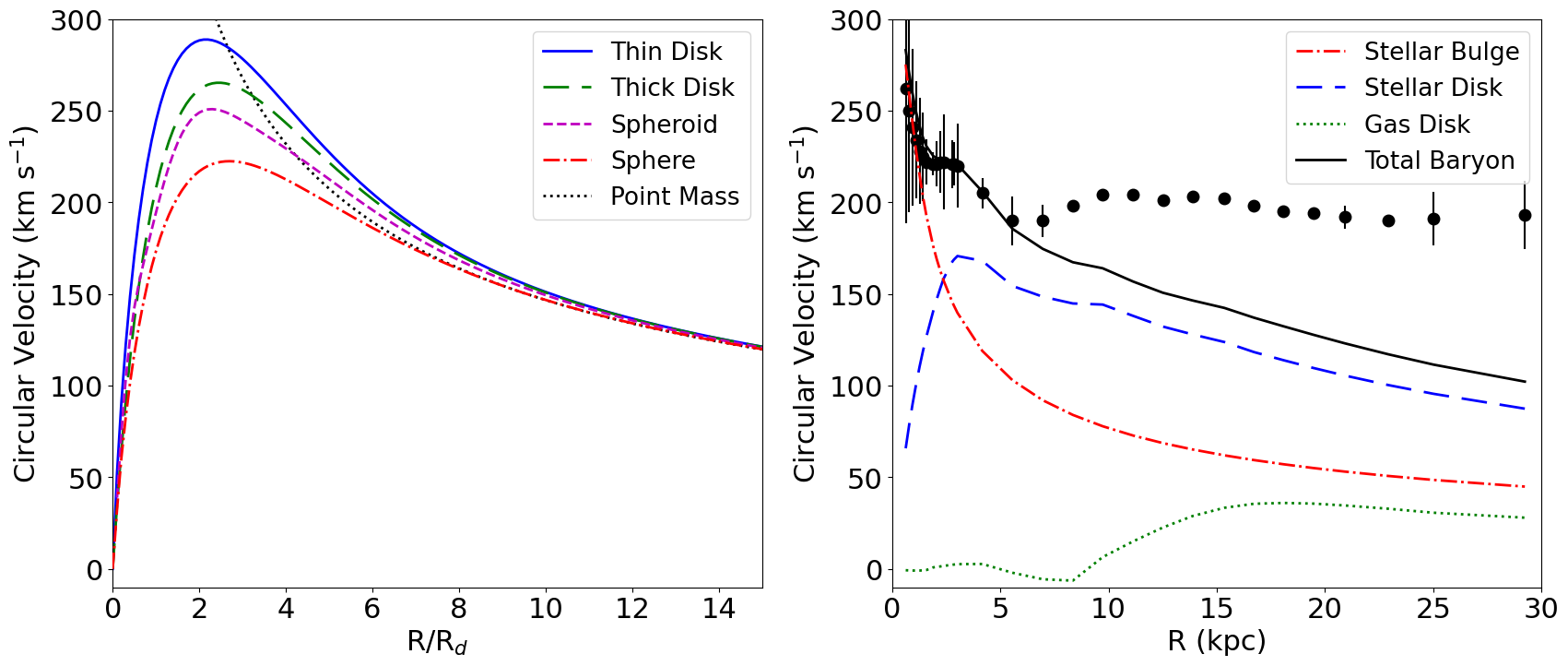}
  \caption{\emph{Left panel}: circular-velocity curves for a model galaxy with a projected exponential density profile and total mass of $5\times10^{10}$ M$_\odot$, assuming different geometries: razor-thin disk (Eq.\,\ref{eq:Freeman}), thick exponential disk with $z_{\rm d}=1/3 R_{\rm d}$ (Eq.\,\ref{eq:Casertano}), oblate spheroid with $q=0.5$ (Eq.\,\ref{eq:Noordermeer} or Eq. \ref{eq:Chandra}), and sphere (Eq.\,\ref{eq:Kent}). The $V_{\rm c}$ of a corresponding point mass is also shown. Flattened mass distributions reach the Keplerian decline at $R\gtrsim 10 R_{\rm d}$. \emph{Right panel}: Mass model for the spiral galaxy UGC\,3546. The H$\alpha$+HI rotation curve (black points with errorbars, \citealt{Noordermeer2007}) is well reproduced at $R\lesssim6$ kpc by the baryonic gravitational field computed from NIR photometry \citep{lelli2016a}. Note that $V_{\rm gas}$ is negative at $5 \lesssim R/{\rm kpc} \lesssim 10$.}
  \label{fig:Vexp}
\end{center}
\end{figure}

In the former approach, in cylindrical symmetry, we have $\rho(R,z)=\Sigma(R)Z(z)$ where $\Sigma(R)$ is the observed radial density profile (from optical/NIR images for stars and emission-line maps for gas) and $Z(z)$ is an assumed vertical density profile. The $Z(z)$ profile can be directly studied only in edge-on disks; common parametrizations are ${\rm sech}^2(z/z_{\rm d})$, $\exp(-z/z_{\rm d})$, or $\exp(-z^2/z_{\rm d}^2)$, where the scale height $z_{\rm d}$ is found to correlate with the scale length $R_{\rm d}$ \citep{vanDerKruit2011}. The velocity contribution at $z=0$ is given by \citep{Casertano1983}:
\begin{equation}\label{eq:Casertano}
V_{\rm_{c}}^{2}(R) = -8 G_{\rm N}R \int_{0}^{\infty} \; \dfrac{\partial\Sigma(\tilde{R})}{\partial\tilde{R}} \left[ \int_{0}^{\infty} Z(\tilde{z})\dfrac{\mathscr{K}(p) - \mathscr{E}(p)}{\sqrt{pR\tilde{R}}} \; d\tilde{z} \; \right] \tilde{R} d\tilde{R},
\end{equation}
where $\mathscr{K}$ and $\mathscr{E}$ are complete elliptic integrals of the first and second kind, respectively, and $p = x - \sqrt{x^{2} - 1}$ with $x = (R^{2} + \tilde{R}^{2} + z^{2})/(2\tilde{R}R)$. Notably, $V_{\rm c}(R)$ depends on $\partial \Sigma(R)/\partial R$ so bumps and wiggles in the mass profile have a relevant effect. In addition, $V_{\rm c}$ at $R$ is given by a double integral in $d\tilde{R}$ and $d\tilde{z}$ from zero to infinity, so depends on the entire mass distribution. This occurs because \emph{Newton's shell theorem does not apply in disks}: mass at $R>R_0$ does contribute to the gravitational field at $R_0$. As a result, $V_{\rm c}^2$ can sometimes be ``negative'' at some radii, in the sense that the gravitational field $V_{\rm c}^2/R$ is directed towards the outer galaxy regions because the mass at $R>R_0$ pulls more than the mass at $R<R_0$. This effect often occurs in the gravitational contribution of \hi\ disks with central holes or strong depressions (see Fig.\,\ref{fig:Vexp}).

If there is a central mass concentration (a ``bulge''), its contribution can be treated separately from the disk. In spherical symmetry, the velocity contribution at $z=0$ is given by \citep{Kent1986}:
\begin{equation}\label{eq:Kent}
V_{\rm_{c}}^{2}(R) =  \dfrac{2\pi G_{\rm N}}{R} \int_{0}^{R} \tilde{R} \Sigma(\tilde{R}) d\tilde{R} + \dfrac{4G_{\rm N}}{R}\int_{R}^{\infty} \left[ \sin^{-1} (R/\tilde{R}) - R(\tilde{R}^2 - R^2)^{-1/2} \right] \tilde{R} \Sigma(\tilde{R}) d\tilde{R}.
\end{equation}
Eq.\,\ref{eq:Kent} is simply $V_{\rm c}^2(r) = G_N M(r)/r$ considering the deprojection of spherical shells of radius $r$.

For an oblate spheroid with intrinsic axial ratio $q$ observed at an inclination $i$, the velocity contribution at $z=0$ is given by \citep{Noordermeer2008}:
\begin{equation}\label{eq:Noordermeer}
V_{\rm_{c}}^{2}(R) = -4 G_{N} \sqrt{q^2\sin^2(i) + \cos^2(i) } \int_{T=0}^{R} \left[ \int_{\tilde{R}=T}^{\infty} \dfrac{\partial\Sigma(\tilde{R})}{\partial{\tilde{R}}} \dfrac{d\tilde{R}}{\sqrt{\tilde{R}^2 - T^2}} \right] \dfrac{T^2 dT}{\sqrt{R^2 - T^2 + q^2 T^2 )}}.
\end{equation}
Assuming that the spheroid's inclination $i$ is the same as that of the gas disk (from kinematic fits, see Sect.\,\ref{sec:rotation}), the mean axial ratio $\tilde{q}$ of the observed isophotes (from optical or NIR images) can be used to infer the intrinsic axial ratio as $q^2 = [\tilde{q}^2 - \cos^2(i)]/\sin^2 (i)$.

For a spheroid where $q$ varies with $R$, the MGE method \citep{Emsellem1994} is most effective. The 2D image is fitted by a sum of $N$ 2D Gaussian functions $j$ with luminosity $L_j$, standard deviation $\sigma_j$, axial ratio $q_j$, and $P.A._{j}$. The potential $\Phi_j(R,z)$ of a Gaussian component stratified on oblate concentric ellipsoids is given by \citep{Chandrasekhar1969, Cappellari2002}:
\begin{equation}\label{eq:Chandra}
\Phi_{j}(R,z) = \dfrac{2G_{\rm N} \Upsilon_j L_j}{\sqrt{2\pi} \sigma_{j}} \int_{0}^{1}
\exp{\left[\dfrac{-T^2}{2\sigma_j^2} \left( R^2 + \dfrac{z^2}{1- T^2 + q_j^2 T^2} \right) \right]} (1-T^2 +q_j^2T^2)^{-1/2} dT
\end{equation}
where $\Upsilon_j$ is the mass-to-light ratio of the component $j$. Thus, one can compute the total potential $\Phi$ and infer the circular velocity $V_c^2(R, z=0) = -R\nabla \Phi$. The MGE approach infers the intrinsic 3D distribution from fitting observed isophotes in the 2D image. This is robust for early-type galaxies with regular isophotes, but becomes more uncertain for late-type galaxies (with spiral arms, star formation, dust lanes, etc.) and for gas maps with irregular distributions.

Figure\,\ref{fig:Vexp} shows $V_{\rm c}(R)$ for a model galaxy with $M=5\times10^{10} M_\odot$ and a projected surface density profile $\Sigma(R)=\Sigma_0 \exp(-R/R_{\rm d})$, using the various equations described above.  In regions probed by gas kinematics ($R\lesssim 6R_{\rm d}-7 R_{\rm d}$), differences in $V_{\rm c}$ due to geometry can be up to $\sim25\%$. Importantly, the Keplerian decline of the corresponding point mass is reached only at $R\gtrsim 10 R_{\rm d}$ or equivalently $R\gtrsim 6 R_{\rm h}$. Thus, the concept of ``dynamical mass'' can be ill-defined unless (1) we model the entire mass distribution for $R\rightarrow \infty$, and/or (2) we have measurements out to very large $R$, where the monopole term of $\Phi(R, z)$ dominates and $M_{\rm dyn}(R) \simeq V_{\rm c}^2 R /G_{\rm N}$.

\subsection{Measuring the baryonic masses}

For a typical galaxy, the expected circular velocity from baryons is given by:
\begin{equation}\label{eq:Vbar}
 V_{\rm bar}^2 = \Upsilon_{\rm bul} |V_{\rm bul}|V_{\rm bul} + \Upsilon_{\rm disk}|V_{\rm disk}|V_{\rm disk} + \Upsilon_{\rm gas} |V_{\rm gas}|V_{\rm gas}
\end{equation}
where $V_{\rm bul}$, $V_{\rm disk}$, and $V_{\rm gas}$ are the contributions from stellar bulge, stellar disk, and gas disk, respectively. $\Upsilon_{\rm bul}$, $\Upsilon_{\rm disk}$, and $\Upsilon_{\rm gas}$ are dimensionless factors (mass-to-light ratios) that scale the contributions according to the total mass of each component, defined for $R\rightarrow\infty$. These factors can be either fixed to some fiducial values, or used as free parameters when fitting the rotation curve, ideally imposing a Gaussian prior in a Bayesian context \citep[e.g.,][]{Li2020}.

The stellar contribution is best traced using NIR images. For late-type disks (Sc-dI), the bulge contribution is often negligible and inner mass concentrations (``pseudobulges'') can be modeled together with the disk. For early-type disks (S0-Sb), instead, it is appropriate to separate bulge \& disk components because they have different geometries and stellar populations. The corresponding mass-to-light ratios can be estimated in several ways \citep[e.g.,][]{McGaugh2015}. Stellar population models typically give $\Upsilon_{\rm bul}\simeq0.7-1.0$ and $\Upsilon_{\rm disk}\simeq0.3-0.7$ at Spizer 3.6 $\mu$m, depending on the assumed star-formation history and chemical enrichment \citep{Schombert2022}. For galaxies on the star-forming main sequence, it is sensible to assume $\Upsilon_{\rm disk}\simeq0.5$ with a 1$\sigma$ variation of $\sim$25$\%$ \citep{McGaugh2015}. Naturally, $\Upsilon_{\rm disk}$ is expected to vary with radius because of metallicity and stellar population gradients, but the effect is somewhat degenerate with geometry (Fig.\,\ref{fig:Vexp}) and subdominant with respect to the absolute calibration in $\Upsilon_{\rm disk}$. Notably, to implement a radially variable mass-to-light ratio, we cannot simply use some $\Upsilon_{\rm disk}(R)$ in Eq.\,\ref{eq:Vbar} but have to recalculate $V_{\rm disk}$ using the integrals in Sect.\,\ref{sec:math} for a rescaled surface density profiles $\Upsilon_{\rm disk}(R)\Sigma_{\rm disk}(R)$. 

The gas contribution is usually dominated by atomic gas, so \hi\ maps are used to compute $V_{\rm gas}$. The conversion from \hi\ luminosity to \hi\ mass is known from atomic physics, so $\Upsilon_{\rm gas}$ must only account for Helium and heavier elements. Considering big bang nucleosynthesis and stellar chemical enrichment, one has $\Upsilon_{\rm gas}\simeq1.34-1.41$ depending on the gas metallicity \citep{McGaugh2020}. The uncertainty on $V_{\rm gas}$ is dominated by the absolute \hi\ flux calibration, which is typically $\sim$10$\%$. The smaller contribution of molecular gas can be separately computed using CO maps \citep{Frank2016}, but CO \& \hi\ data are rarely available for the same galaxy samples. Luckily, molecular gas is distributed in a similar way as the stellar disk (similar $R_{\rm d}$), so its contribution in star-forming galaxies can be roughly taken into account with a systematic correction to $V_{\rm disk}^2$ of the order of 7$\%$ \citep{McGaugh2020}, or simply included in the error budget of $\Upsilon_{\rm disk}$. Warm ionized gas ($T\simeq10^4$ K) typically gives a negligible mass contribution, while hot ionized gas ($T\simeq10^{6-7}$ K) is thought to form low-density halos extending over hundreds of kpc, so its gravitational contribution is negligible within the \hi\ disk.

\begin{figure}[t]
\begin{center}
  \centerline{\vbox to 6pc{\hbox to 10pc{}}}
  \includegraphics[width=\textwidth]{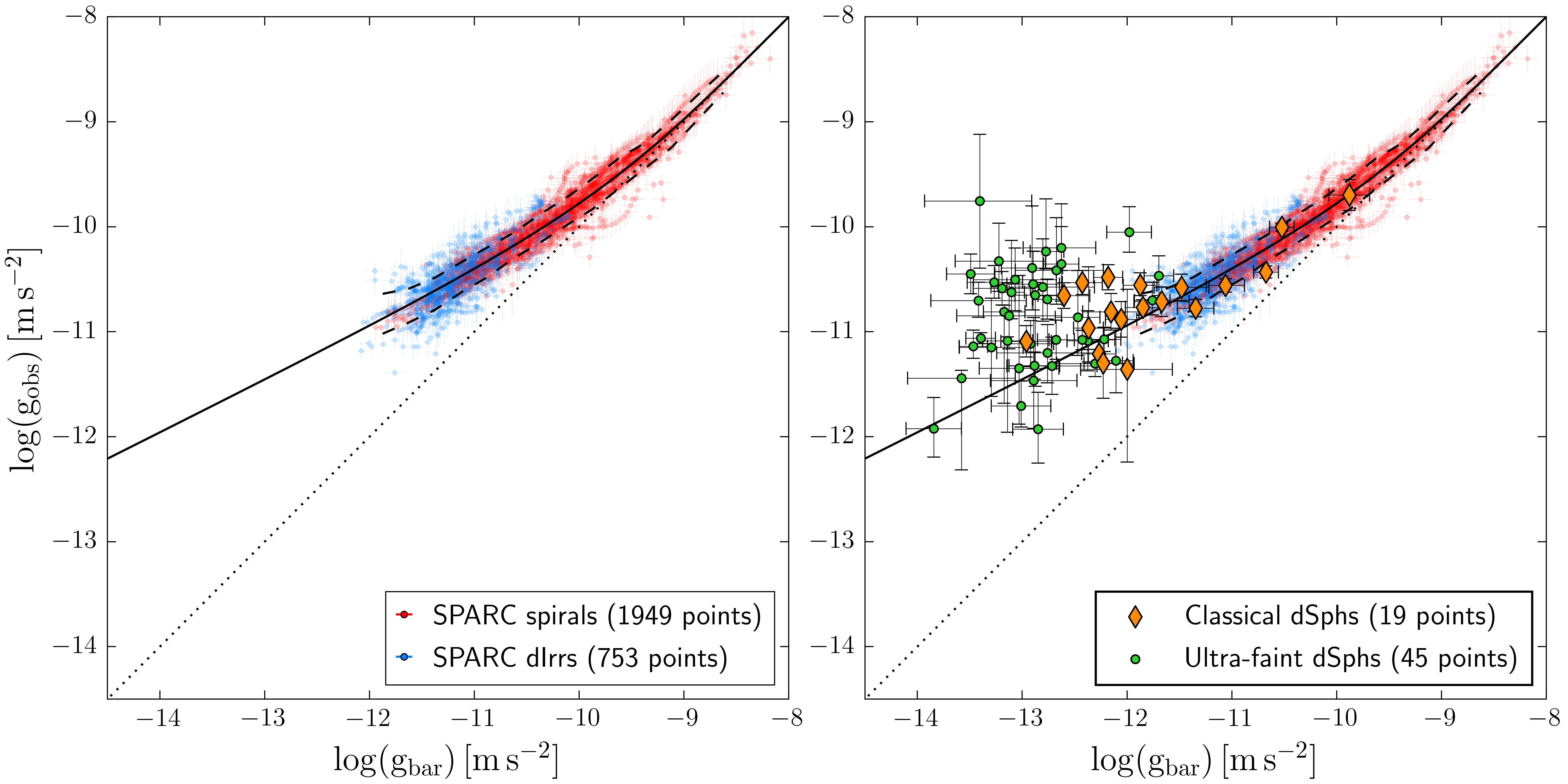}
  \caption{The Radial Acceleration Relation (RAR): the observed centripetal acceleration at each radius is plotted against that expected from the baryons' distribution. The left panel shows the RAR from 153 disk galaxies from the SPARC database; blue and red dots correspond to data from spirals and dIrrs, respectively. The right panel adds 62 dSphs in the LG; orange diamonds and green circles show ``classical'' dSphs and ultra-faint dSphs. In both panels, the dotted line is the line of unity; the solid line shows a fit to the SPARC data; dashed lines correspond to 1 standard deviation from the mean. Adapted from \citet{Lelli2017}.}
  \label{fig:rar}
\end{center}
\end{figure}
The DM contribution can be added to Eq.\,\ref{eq:Vbar} assuming a spherical halo with a given volume density profile. Equations for common halo models are summarized in \citet{Li2020}. 
%For self-consistent mass models in which the DM halo responds to the baryonic gravitational field and adiabatically contracts, we refer to \citet{Li2022}.

\section{Radial acceleration relation}\label{sec:rar}

Thanks to large \hi\ and NIR surveys, mass models have been built for hundreds of nearby galaxies, spanning over $\sim$5 dex in stellar mass and $\sim$3 dex in effective surface brightness \citep[e.g.,][]{lelli2016a, Iorio2017}. The scientific implications of these mass models have been recently reviewed in \citet{McGaugh2020} and \citet{Lelli2022}. Here we only recall some basic facts. In high-mass ($M_\star > 3 \times 10^{9}$ M$_\odot$) and high-surface-brightness (HSB) galaxies, baryonic matter can generally explain the inner dynamics ($R\lesssim1-2 R_{\rm h}$) while the DM effect appears in the outer regions. In low-mass ($M_\star \leq 3 \times 10^{9}$ M$_\odot$) and low-surface-brightness (LSB) galaxies, instead, the DM effect is already important at small radii. In addition, the baryonic distribution and the rotation-curve shapes appear tightly coupled in galaxies at a local level \citep{Sancisi2004}.

The RAR (Fig.\,\ref{fig:rar}) is an effective way to quantify the \emph{local} baryon-dynamics coupling in galaxies \citep{McGaugh2016, Lelli2017}. At each radii, the observed acceleration from rotation curves ($g_{\rm obs}=V_{\rm rot}^2/R$) correlates with that expected from the distribution of baryons ($g_{\rm bar} = V_{\rm bar}^2/R = -\nabla \Phi_{\rm bar}$). At high accelerations, $g_{\rm obs}=g_{\rm bar}$ so there is no need of DM. At low accelerations, below a characteristic acceleration scale of $\sim$10$^{-10}$ m s$^{-2}$, $g_{\rm obs}>g_{\rm bar}$ and the DM effect emerges. Intriguingly, in the RAR plane, the outer DM-dominated regions of spiral galaxies smoothly overlap with the inner DM-dominated regions of dwarf irregulars (dIrrs) as if the two regions ``know'' about each other. The observed scatter of the RAR is just $\sim25\%-30\%$, so its intrinsic scatter must be tiny, if not zero \citep{Li2018, Desmond2023}.

Remarkably, the existence and properties of the RAR were predicted \emph{a-priori} by \citet{Milgrom1983b} using his Modified Newtonian Dynamics (MOND). It is thus important to test whether different galaxy types follow the same RAR. Early-type galaxies (ellipticals and lenticulars) lie on the same RAR of spirals and dIrrs \citep{Lelli2017, Shelest2020}. The situation is more uncertain for dwarf spheroidal galaxies (dSphs) in the LG. Given their low surface densities, dSphs are truly unique systems to study the low RAR down to very low accelerations, but they lack a rotating gas disk, so $g_{\rm obs}$ must be inferred from stellar kinematics. 

\citet{Lelli2017} compiled a sample of 62 dSphs with stellar velocity dispersion $\sigma_\star$ from single-star spectroscopy. Assuming that dSphs are fully pressure-supported and spherically symmetric, we can compute $g_{\rm obs} = 3 \sigma_\star^2 / r_{1/2}$ and $g_{\rm bar} = G_{\rm N} M_\star / (2 r_{1/2}^2)$, where $r_{1/2}$ is the 3D half-light radius. This choice is motivated by the fact that the mass-anisotropy degeneracy in spherical, pressure-supported systems is nearly broken at $r_{1/2}$ \citep{Wolf2010}. For dSphs, therefore, we have a single point per galaxy rather than a spatially resolved analysis. Figure\,\ref{fig:rar} (right panel) show that ``classical'' dSphs roughly follow the same RAR as disk galaxies, whereas ultra-faint dwarfs (UFDs) display a much larger scatter and are systematically shifted towards higher $g_{\rm obs}$ than expected from the RAR extrapolation. Several observational uncertainties may affect the location of UFDs on the RAR: (1) $M_\star$ and $r_{1/2}$ are not inferred from NIR surface photometry but from star counts in optical images, after candidate stars are selected using color-magnitude diagrams and template isochrones, so there could be systematics between different datasets, (2) $\sigma_\star$ may be inflated by undetected stellar binaries and/or small number statistics, especially when only a dozen of bright stars are available, and (3) both $\sigma_{\star}$ and $r_{1/2}$ may be inflated by tidal forces from the host galaxies (the Milky Way and Andromeda), possibly driving the systems out of dynamical equilibrium. Major observational and theoretical efforts are truly needed to clarify the location of UFDs on the RAR.

\vspace{-0.2cm}
\begin{discussion}

\discuss{Marina Rejkuba}{What are the prospects to use the stellar $V_{\rm rot}$ in low surface brightness outskirts? Maybe in resolved galaxies?}

\discuss{Federico Lelli}{It is hard to study stellar kinematics beyond 1-2 $R_{\rm h}$ with existing integral-field spectrographs due to the low surface brightness. Resolved individual-star spectroscopy is currently limited to LG galaxies and require large spectrophotometric campaigns. Future 40-meters telescopes, such as the ESO ELT, will surely push stellar kinematics further. However, in addition to the observational challenges, there is a theoretical limitation. To infer the circular velocity tracing the gravitational field, we need to correct the stellar $V_{\rm rot}$ for pressure support, which requires assumptions on the shape of the velocity dispersion tensor.}

\discuss{Marina Rejkuba}{You brought up uncertainties in $R_{\rm h}$ as possible explanation for the scatter in the RAR for LG dwarfs. What did you use for $R_{\rm h}$? In particular, for ultra-faint dwarfs, this can indeed be uncertain as they may be out of equilibrium or affected by the Milky Way.}

\discuss{Federico Lelli}{We compiled $R_{\rm h}$ from various literature sources, then we calculated the deprojected 3D half-light radius $r_{1/2}$ that is needed to compute centripetal accelerations. In addition to the technical challenges in measuring $R_{\rm h}$ in ultra-faint dwarfs, I fully agree that out-of-equilibrium dynamics is a major concern because it could artificially increase both $R_{\rm h}$ and the measured velocity dispersion $\sigma_{\star}$. As far as I can tell, the observed scatter in the RAR may be entirely driven by observational uncertainties.}

\discuss{Yabin Yang}{About the dSphs scattered distribution on the RAR, perhaps taking their distances to be an extra parameter could help to understand their large dispersion?}

\discuss{Federico Lelli}{We used the distances of dSphs from their host galaxies (Milky Way or Andromeda) to compute the expected gravitational acceleration ($g_{\rm host}$) and tidal acceleration ($g_{\rm tides}$) at their location. The scatter in the RAR decreases imposing a cut in $g_{\rm obs}/g_{\rm tides}$, but there is no simple trend, nor obvious cut to use. The observed scatter is probably driven by multiple effects and uncertainties, so considering a single quantity does not tell the full story.}

\discuss{Ting Li}{What should the velocity dispersion be for the ultra-faint dwarfs if they follow the same RAR as disk galaxies?}

\discuss{Federico Lelli}{On average, the velocity dispersion of ultra-faint dwarfs should be a factor of 2 smaller to have them on the low-acceleration extrapolation of the RAR of disk galaxies.}

\end{discussion}

\vspace{-0.2cm}
\bibliography{GasDynamics}
\bibliographystyle{iaulike}

\end{document}